# The outreach activities in the astronomical research institutions and the role of librarians: what happens in Italy*

*by Monica Marra (INAF – Italian National Institute of Astrophysics)*
email: monica.marra@oabo.inaf.it

**Abstract**

The outreach activities can be considered a new frontier of all the main astronomical research institutions worldwide and are a part of their mission that earns great appreciation from the general public. Here the situation at INAF, the Italian National Institute for Astrophysics, is examined and a more active role for librarians is proposed.

**1. Introduction**

It is not surprising nowadays that among the official missions of the world's main research institutions in astronomy and related fields you find the outreach activities. This is true for example for NASA, which indicates as its third mission that of "inspiring the next generation of explorers" and of "engaging the public in shaping and sharing the experience of exploration and discovery" [1]. NASA is certainly very serious in this, as its FY 2004 budget request for education is 170 million dollars [2].
Although we can't say how much it spends for its outreach initiatives, we do know that NASA's European "colleague", ESA (European Space Agency), has an education office as well, which is carrying on a number of interesting projects for all study levels - with a very good outcome on the web at http://www.esa.int/export/esaED/index.html. Recently, in 2001 the European Southern Observatory - the well-known intergovernmental European organization born in 1962 for astronomical research - established a dedicated Educational Office as a part of its Education and public relations department [3]. An Outreach network with 12 member scientists or communication specialists from different countries was also provided [4].
In Italy, the main research institution in astronomy and astrophysics is now INAF (National Institute for Astrophysics) [5]. INAF was born with a national law dated 1999, by merging the twelve, formerly independent astronomical observatories located in Italy [6]. From the very first beginning, one of the four missions of INAF, as indicated on the law that gave birth to it, was that of "promoting the astronomical knowledge in school and society, also by means of appropriate outreach and museum activities". The further, recent national law that modified INAF's organization hasn't changed much under this point of view, although missions may perhaps have been less clearly stated [7]: actually, one of the first national initiatives that have been started after the launch of the "new INAF" is a national meeting on interactions among this institution and the media.
In fact, the outreach activities (mainly, conferences and public observations), were a tradition in the Italian astronomical observatories - as well as in most other countries - even before the birth of INAF, but it is widely recognized also by the present Institution itself that they generally used to be scarcely organic and often due to astronomers' individual initiatives, to which, moreover, some professional jealousy and resistance to better coordination was not always completely foreign. The birth of a national institution can of course make a difference for as much as this matter is concerned.
The publication last year of the first INAF triennial plan for the years 2003-2005 and the related official papers [8] enable us to get a clear idea of this institution's goals for the next future and also to have a first, important overview of the most relevant national initiatives in all INAF's fields of activity. This includes also the outreach and educational activities, to which a whole chapter is devoted in the main document, and another, even more detailed, in the report written by the Department for Observatories.

## 2. The Italian National Institute for Astrophysics (INAF) and its outreach activities: a story of success

The expenses made by INAF in 2002 for the "diffusion of scientific culture, outreach and information" amount to 200 kEuro. In the years 2003-2005 they will increase by 50%, with an expense of 300 kEuros per year. In a period that is internationally known for being critical under an economic point of view, this is a further proof of the strategic role that INAF acknowledges to the external conveyance of its results in research.

Just to detail about the number of people who have visited the INAF centers during public openings and conferences, we are informed that almost 200.000 visits were made: roughly, we can calculate an average of more than 16.600 interested people for each of the 12 centers. But this is not enough: "the demand of qualified outreach activities in the field of the scientific culture is higher that the possibility of supply by the astronomical observatories", as the same organism seems to be sorry to say.

An exact idea of the activities to which INAF recognizes a nature of excellence within its outreach ones is given by the list provided in the paper by the Department for Observatories, mentioned above [9]. By type, they are about 20 from all the former Observatories, now INAF centers. Two initiatives are partly included in the traditional ones I've already mentioned above, e.g. visits to the telescopes and Observatories' web sites: all the centers, with no exception, provide them. An examination of the list shows that some particular centers concentrate a higher number of these best-recognized activities: Padua, Bologna, Rome, Catania, Brera are top of the list, with 6,5,4 specially good initiatives each. Particularly noteworthy, according to me, is the number of very good web sites (also award-winners) that the Padua center has devoted to astronomy for children [10]. Other mentions of honor were got by the seven astronomical museums (Bologna, Brera, Naples, Padua, Palermo, Rome, Trieste); by three planetariums (Arcetri, Naples and the virtual one in Padua) and one interactive program on the same subject (Catania); by some refresher courses for astronomy teachers (Bologna and others); by the first "national olympic games of astronomy" (Trieste); by the continuing didactic exhibition Astrolab and the project DivA (Rome); by the miniature solar system model (Bologna); by the multimedia room for the public in Asiago (Padua); by some itinerant exhibitions held in Brera, Catania, Bologna, sometimes in collaboration with other important institutions in this field such as ASI (Italian Space Agency).

## 3. When success is "too much": why not employ librarians?

It is generally acknowledged and is no doubt that astronomy and astrophysics, for their very "hard" scientific core, are strongly bound to the scientists who deal with it. On the other side, an amount of activities for the public that is increasing for number and type, and the institutional will to make it grow furtherly, do require an evaluation of the amount and type of human resources to devote to non-research activities. It is interesting to see what the situation is at present, which we can do thanks to the Observatories' Annual Reports for the year 2002 [11]. With the exception of one center alone, and despite a partition of the reports which is not the same for all, they all give account also of the outreach activities, and almost always detail about the local staff committed to the different projects.

The core of the staff that projects and carries out these activities is practically always composed of astronomers. Things may change when these activities are run-in, e.g. for public observations, in which case observation technicians may have a leading role and be charged to carry out the sessions completely. Another task for which a support from non-researchers is necessary is the custody of the premises during extra openings, something that involves responsability due to the value of the scientific instruments and equipment, and external contacts. The first has seen an involvement by external working forces, as it is said for Palermo (5 social workers from the local authorities help with the extra openings) and in a way for Rome (15 physics university students collaborate as guides for the public).

As for the second, the existence of scientific secretariats seems to be something informal and situation-dependent in all centers, something we can more easily understand if we consider their small dimension (from a minimum of 15 to a maximum of 53 technical, administrative and general support staff in the 12 centers), but anyway is far from seeming ideal.

New activities, instead, seem to be in full control of the researchers. This seems to be absolutely correct for as much as the contents are concerned; but if we assume that the *mode* of these contents' conveyance to the public is of some importance, and that the public expectation should be known before starting planning, we should conclude that broader teams should be provided. People in the staff with experienced communication skills, or experience in specific sectors of the public (children, for instance) should have the possibility to contribute actively in the conception itself of new initiatives, and not only be involved, even when it happens, in their performance.

The present trend in the INAF centers seem to be exploring two main different options. Outsorcing the information and outreach activities seems to be one of the best-liked ways to find human resources for these activities: this happens in Arcetri (Florence), with its volunteer Committee for the Divulgation of Astrophysics, Palermo, with the social workers mentioned above, Rome (university students) and Turin, which employs an

Association for the Divulgation of Astrophysics. In all these cases a consideration appears to have been done about the low-cost of this kind of cooperators.

Other interesting approaches to the problem have been tried by the centers of Cagliari, Brera and Rome, which have settled specific offices or stable working groups with dedicated internal personnel. In fact, the very professional staff of these offices is composed by astronomers, as in the DivA working group in Rome, or by people who have an astronomy degree and have then chosen a technical career, as in Brera, or by a technician alone, as in Cagliari.

Librarians have much to offer about this. As information professionals, they must be able to communicate with their users for satisfying their information needs effectively: and they do so daily, by using the most up-to date technologies. Updating online catalogues, doing internet searches and in some cases (Bologna, Turin) designing and updating their libraries' web pages on themselves is daily routine for them. The average age of these 24 professional is around 40, which enables them to be both reliable and able to learn easily again new technologies and approaches to the matter; moreover, the great majority of them holds a university degree. At present, nevertheless, only the center in Naples seems to dedicate its librarians also to the center's educational activities, with good results besides.

The real obstacle to a better employment of this educated and no-additional-cost personnel with regards to the outreach activities, however it may be conceived, seems to be two-fold. On one side, librarians are often perceived by scientists as "humanists" and in a way seem to suffer for a sort of reversed cultural suspicion, when activities connected with the astronomical research are concerned. On the other side, it has often been assumed that an appropriate involvement of librarians in the centers' educational or outreach initiatives should anyway have to do with older documents, e.g. exhibitions of old books or historical archives. In fact, the main ability of a librarian should be that of understanding and fulfilling the information needs of his/her present time: a new approach to the matter, after the birth of the new national institution, should be attempted.

**References**

[1] See NASA's strategic plan for the year 2003 at <http://www.nasa.gov/pdf/1968main_strategi.pdf>
[2] See NASA's short report on its budget for education programs at http://www.nasa.gov/pdf/1973main_edprograms1.pdf
[3] ESO Educational office can exhibit a rich website at <http://www.eso.org/outreach/eduoff/>; at http://www.eso.org/outreach/> a variety of ESO outreach activities and materials are shown.
[4] See at http://www.eso.org/outreach/epr/net-contact.html (update: December 2002 at our visit on 29.9.2003).
[5] INAF official website has an English version of the home-page at <http://www.inaf.it/english/english_home.html>
[6] The law mentioned above is exactly D.Lgs. n.296, dated 23.07.1999. The twelve observatories are located in: Arcetri (Florence), Bologna, Brera (Milan), Cagliari, Catania, Capodimonte (Naples), Palermo, Padua, Rome, Teramo, Turin, Trieste. Individual web sites and physical locations at <http://www.inaf.it/strutturediricerca.htm>
[7] D.Lgs. n.138, dated 4.6.2003.
[8] See all them (in Italian only) at http://www.inaf.it/pianotriennale/pianotriennale.htm
[9] See chapter 6.
[10] Some of them are available in English at <http://www.pd.astro.it/public/publicen.html>
[11] The individual urls of the twelve centers' annual reports have been grouped at <http://www.inaf.it/annualreports.htm>